\documentclass[epj]{svjour}
\usepackage{latexsym}
\usepackage{graphics}
\usepackage{graphicx}
\usepackage{amssymb}   
\usepackage{url}
\usepackage{amsmath}
\usepackage{aas_macros}
\begin{document}
\title{Constraining r-process nucleosynthesis via enhanced accuracy neutron-capture experiments}
\subtitle{Post LS3 prospects at CERN n\_TOF}
%
%
%
\date{Received: date / Revised version: date}
%
\authorrunning{C. Domingo-Pardo et al.}

\author{
C.~Domingo-Pardo$^{1}$\footnote{email: domingo@ific.uv.es}, %
C.~Lederer-Woods$^{2}$, %
A.~Mengoni$^{3,4,5}$
}
\institute{
$^{1}$Instituto de F\'{\i}sica Corpuscular, CSIC - Universitat de Val\`encia, Spain \newline
$^{2}$School of Physics and Astronomy, University of Edinburgh, United Kingdom\newline
$^{3}$Agenzia nazionale per le nuove tecnologie, l'energia e lo sviluppo economico sostenibile (ENEA), Italy\newline
$^{4}$Istituto Nazionale di Fisica Nucleare, Sezione di Bologna, Italy\newline
$^{5}$European Organization for Nuclear Research (CERN), Switzerland\newline 
}


%
%

\date{Received: \today / Revised version: \today}

\abstract{The isotopic abundances of r-process elements in the solar system are traditionally derived as residuals from the subtraction of s-process contributions from total solar abundances. However, the uncertainties in s-process nucleosynthesis—particularly those arising from Maxwellian Averaged Cross Sections (MACS)—propagate directly into the r-process residuals, affecting their reliability. Building upon the seminal work of Goriely (1999), who introduced a multi-event s-process model to quantify these uncertainties, we revisit the problem using a simplified yet effective approach. By assuming that the relative uncertainty in s-process isotopic abundances scales linearly with the MACS uncertainties from data libraries (KADoNiS), we identify a subset of isotopes for which the r-process residuals remain significantly uncertain. Using updated solar abundances (Lodders 2025) and s-process contributions from Bisterzo et al. (2014), we present a short list of isotopes that are prime candidates for improved ($n,\gamma$) measurements at CERN n\_TOF in the near future. Our analysis provides a practical framework for prioritizing future experimental efforts that will profit from upgrades and enhancements of the n\_TOF facility. It also highlights the need to revisit key neutron-capture cross sections to refine our understanding of the r-process isotopic abundance pattern, commonly used as a benchmark in stellar models of explosive nucleosynthesis.}
\PACS{
      {PACS-key}{discribing text of that key}   \and
      {PACS-key}{discribing text of that key}
     } 
%
\maketitle
\section{Introduction}
Understanding the origin of elements heavier than iron remains one of the fundamental challenges in nuclear astrophysics. These elements are synthesized primarily through two neutron-capture processes: the slow neutron-capture process (s-process) and the rapid neutron-capture process (r-process). The s-process comprises two components: the weak s-process, occurring in massive stars (M $>$ 8 M$_{\odot}$) during core He and shell C burning and responsible for production in the A $\approx$ 60-90 mass range \cite{Pignatari10}, and the main s-process, occurring in low- and intermediate-mass AGB stars (1M$_{\odot}\le $M $\le$ 3M$_{\odot}$) and responsible for production in the A = 90-200 range\,\cite{Gallino98,Kaeppeler11,Lugaro23}. The r-process takes place in explosive environments such as neutron star mergers and core-collapse supernovae\,\cite{Arcones23}.


The seminal work of Goriely \cite{Goriely99} introduced a multi-event s-process model to quantify how uncertainties in nuclear reaction rates propagate into s-process abundance predictions and, consequently, into r-process residuals. Building on this framework and incorporating the updated s-process calculations of \cite{Bisterzo14}, recent galactic chemical evolution models have provided refined estimates of s-process contributions across the nuclear chart. Nevertheless, significant uncertainties remain for several key isotopes, where large MACS uncertainties translate directly into poorly constrained r-process abundances. These uncertainties limit our ability to test theoretical models of r-process nucleosynthesis and to understand the astrophysical sites responsible for the production of the heaviest elements.

Over two decades, the CERN n\_TOF facility has demonstrated its capability to measure neutron capture cross sections with high accuracy across the energy range relevant for stellar nucleosynthesis. After the CERN Long Shutdown 3 (LS3), which includes various facility upgrades, n\_TOF will offer enhanced opportunities for measurements of isotopes that are currently limiting the precision of r-process abundance determinations. This paper identifies a prioritized list of isotopes for which improved $(n,\gamma)$ measurements at n\_TOF would significantly reduce uncertainties in r-process residuals.

Although the primary focus of this work is on improving the accuracy of r-process isotopic abundances, we also discuss relevant aspects of s-process nucleosynthesis for each selected isotope. Many of these isotopes are located at critical positions along the s-process path---such as branching points or neutron magic numbers---where accurate cross section measurements not only reduce r-process uncertainties but also provide important constraints on s-process physics, including neutron densities, temperatures, and mixing processes in AGB stars. 

From a technical perspective, the neutron-capture measurements at n\_TOF employ 
C$_6$D$_6$ liquid scintillation detectors in combination with the Pulse Height 
Weighting Technique (PHWT) \cite{Moxon63,Macklin67}. This method, which requires low-efficiency $\gamma$-ray 
detectors and detailed Monte Carlo simulations of the experimental setup to derive 
energy-dependent weighting functions, has been extensively validated at n\_TOF. 
Dedicated systematic studies \cite{Abbondanno04,Domingo04} have demonstrated that the PHWT, when 
applied with detailed Monte Carlo (MC)-based weighting functions tailored to each specific sample-detector 
configuration, can achieve reaction yield determinations with an accuracy of 2\% or 
better. This validation, performed using samples of various dimensions and compositions, 
confirmed that the technique properly accounts for the influence of sample geometry 
and surrounding materials on the detector response, which is a critical requirement for 
high-precision cross-section measurements. These developments now enable systematic 
cross-section determinations at n\_TOF with enhanced accuracy, directly addressing the 
need for improved nuclear data to constrain r-process residuals and refining s-process models of nucleosynthesis \cite{Domingo23b,Domingo25}. 

\section{Methodology}\label{sec:methodology}
Our approach to identifying high-priority isotopes for neutron capture measurements is based on a simplified yet effective propagation of nuclear data uncertainties to r-process residuals. To this aim we utilize three primary data sources:
\begin{itemize}
\item MACS values and uncertainties: The Karlsruhe Astrophysical Database of Nucleosynthesis in Stars (KADoNiS) version 1.0 \cite{kadonisv1}, which provides recommended Max\-wel\-lian-averaged cross sections at $kT = 30$ keV along with their uncertainties for isotopes relevant to s-process nucleosynthesis.
\item Solar system abundances: The updated solar abundance compilation by Lodders et al. (2015) \cite{Lodders2025}.
\item S-process contributions: The galactic chemical evolution calculations of Bisterzo et al. (2014) \cite{Bisterzo14}, which provide predictions for the s-process contribution to each isotope's solar abundance based on detailed AGB stellar models with updated $^{13}$C-pocket structures and neutron capture networks.
\end{itemize}
The r-process residual abundance for a given isotope is obtained by subtracting the s-process contribution from the total solar abundance:
\begin{equation}
N_{\rm r} = N_{\odot} - N_{\rm s}
\end{equation}
where $N_{\odot}$ is the solar system abundance and $N_{\rm s}$ is the s-process contribution predicted by stellar models.

\begin{figure*}[!htbp]
  \centering
  \includegraphics[width=0.96\textwidth]{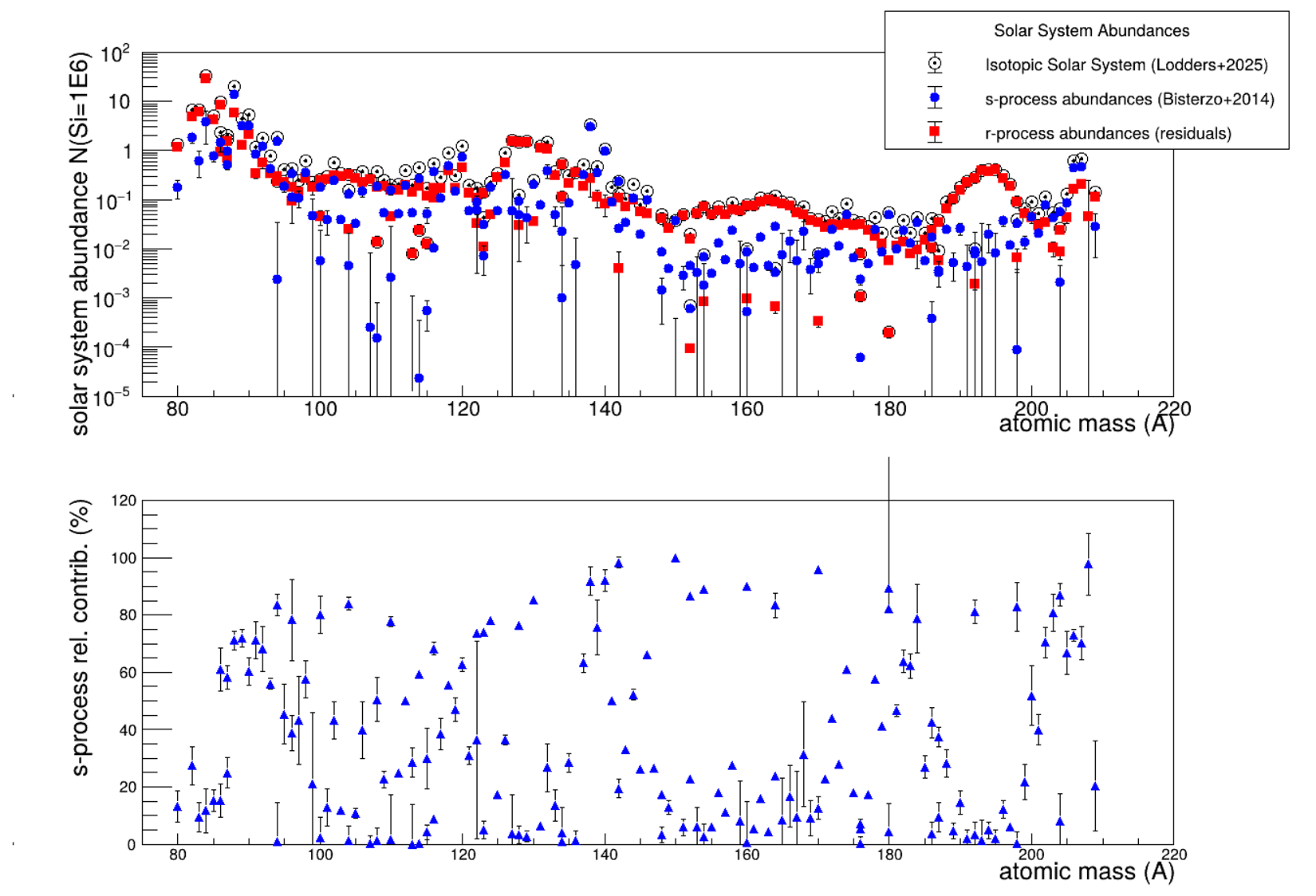}
  \caption{Isotopic abundances in the solar system (normalized to Si $= 10^6$) showing the decomposition into s-process and r-process components. Upper panel: Total solar system abundances from \cite{Lodders2025} (open circles), s-process contributions from \cite{Bisterzo14} (blue filled circles), and derived r-process residuals $N_{\rm r} = N_{\odot} - N_{\rm s}$ (red filled circles). Lower panel: Relative s-process contributions (percentage of total solar abundance). Error bars on r-process residuals reflect the propagated uncertainties from MACS values as described in Section~\ref{sec:methodology}.}
\label{fig:abundances}
\end{figure*}

To estimate the uncertainty in the r-process residual, we employ a linear propagation approximation. We assume that the relative uncertainty in the s-process abundance for a given isotope scales linearly with the relative uncertainty in its MACS:
\begin{equation}
\frac{\Delta N_{\rm s}}{N_{\rm s}} \approx \frac{\Delta \sigma}{\langle \sigma \rangle}
\end{equation}
where $\Delta \sigma / \langle \sigma \rangle$ is the relative MACS uncertainty from KADoNiS 1.0. Although this approximation neglects second-order effects and correlations between isotopes in network calculations, it provides a practical first-order estimate for identifying the most problematic cases.

The approximation is well justified by the s-process equilibrium conditions in AGB stars: in the mass range $A = 90$--200, the product $N_s \langle \sigma \rangle$ remains approximately constant \cite{Kaeppeler11}, which directly supports the assumed linear scaling between MACS uncertainties and s-process abundance uncertainties. We note that our analysis focuses exclusively on the main s-process component produced in AGB stars. The weak s-process component, follows different nucleosynthesis conditions \cite{Pignatari10} and is not addressed in the current work. Nevertheless, the methodology presented here could be extended to that mass region in future studies, where uncertainties in r-process residuals are expected to be similar or even more significant.

Given that uncertainties in solar system abundances from \cite{Lodders2025} are typically small ($\sim$1--3\%) compared to the dominant MACS uncertainties (ranging from $\sim$10\% to $>$50\% for many isotopes), we approximate the uncertainty in the r-process residual as:
\begin{equation}\label{eq:DeltaNr}
\Delta N_{\rm r} \approx \Delta N_{\rm s} = N_{\rm s} \cdot \frac{\Delta \sigma}{\langle \sigma \rangle}
\end{equation}

This simplified approach provides a practical lower limit on the uncertainty in r-process residuals, as it assumes MACS uncertainties dominate over other potential error sources such as uncertainties in s-process stellar models, beta-decay rates at branching points, and temperature-dependent effects. While this assumption is well justified for many isotopes where MACS uncertainties exceed 10--15\% and solar abundance uncertainties remain below 3\%, there are cases, particularly at certain branching points, where additional uncertainty contributions become significant (see discussion of $^{99}$Ru in Section \ref{sec:cases}). Nevertheless, this first-order approximation successfully identifies isotopes where improved MACS measurements would have substantial impact on constraining r-process yields. A more comprehensive uncertainty analysis incorporating all error sources through full s-process network calculations would be valuable but is beyond the scope of this prioritization study.



For this study we identify isotopes as high-priority targets for improved $(n,\gamma)$ measurements based on two criteria:
\begin{enumerate}
\item The relative uncertainty in the MACS from KADoNiS 1.0 must be $\geq 10\%$
\item The resulting relative uncertainty in the r-process residual must be $\geq 20\%$
\end{enumerate}

The first criterion ensures that the isotope's MACS is not already well-constrained by existing measurements, while the second criterion guarantees that the uncertainty propagates significantly into the r-process abundance determination. Isotopes meeting both criteria represent cases where improved cross section measurements would have the greatest impact on refining our understanding of r-process nucleosynthesis.

Table~\ref{tab:isotopes} lists the isotopes satisfying these criteria, showing their r-process residual abundances ($N_{\rm r}$), absolute uncertainties ($\Delta N_{\rm r}$), and the relative uncertainties in both the r-process residuals and the MACS values. In the following sections, we discuss each of these isotopes in detail, examining the current status of their cross section measurements, the astrophysical implications of the uncertainties, and the prospects for improved measurements at the CERN n\_TOF facility.
\begin{table}[h]
\centering
\begin{tabular}{lcccc}
\hline
Isotope & $N_r$ (Si=$10^6$) & $\Delta$N$_r$/N$_r$(\%) & $\Delta \sigma/\langle \sigma \rangle$ (\%) \\
\hline
$^{97}$Mo & $1.4(4) \times 10^{-1}$ & 27 & 15 \\
$^{99}$Ru & $1.8(6) \times 10^{-1}$ & 32 & 25 \\
$^{122}$Sn & $1.1(6) \times 10^{-1}$ & 54 & 35 \\
$^{139}$La & $1.1(5) \times 10^{-1}$ & 39 & 10 \\
$^{168}$Er & $4.9(13) \times 10^{-2}$ & 27 & 18 \\
$^{184}$W & $9.3(53) \times 10^{-3}$ & 57 & 12 \\
$^{200}$Hg & $4.3(9) \times 10^{-2}$ & 22 & 10 \\
\hline
\end{tabular}
\caption{Isotope data with Nr values, uncertainties, and relative uncertainties for $N_r$ and MACS.}
\label{tab:isotopes}
\end{table}

\section{Case Studies of Key Isotopes}\label{sec:cases}

\subsection*{$^{97}$Mo}
\begin{figure}[!htbp]
  \centering
  \includegraphics[width=0.95\columnwidth]{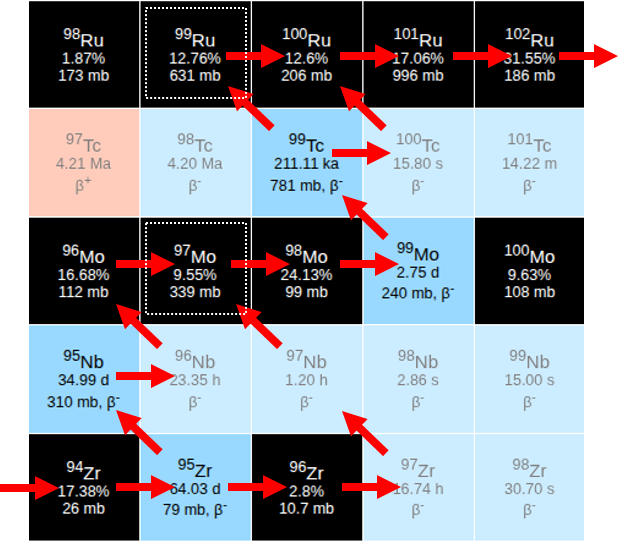}
  \caption{Nuclear chart segment showing the $^{99}$Tc branching point that determines $^{99}$Ru production in AGB stars.}
\label{fig:MoRu}
\end{figure}

Molybdenum isotopes serve as critical tracers of s-process nucleosynthesis in presolar silicon carbide grains from AGB stars \cite{Stephan2019}. While $^{96}$Mo is especially valuable as an s-process-only isotope (see Fig.\ref{fig:MoRu}), shielded from r-process production by $^{96}$Zr, $^{97}$Mo is a mixed s- and r-process isotope contributing about 47\% to the solar inventory via s-process. The high-precision measurements of these isotopes in mainstream SiC grains \cite{Liu24} reveal that their compositions follow clear mixing lines between a pure s-process progenitor and solar material, with $^{96}$Mo and $^{97}$Mo abundances closely following the inverse of their neutron-capture cross sections. This behavior shows remarkably constant s-process ratios across different SiC grains, indicating that the branching at $^{95}$Zr remains minimal even when the $^{22}$Ne($\alpha$,n)$^{25}$Mg reaction is marginally activated during thermal pulses at peak neutron densities of $\sim10^{10}$ cm$^{-3}$ \cite{Stephan2019,Lugaro03,Liu2014}. This weak branching is sufficient to produce the observed $^{96}$Zr abundances without significantly bypassing $^{96}$Mo in favor of $^{97}$Mo and $^{98}$Mo. The observed variations in other Mo isotopes like $^{94}$Mo, $^{95}$Mo, and $^{100}$Mo provide unique constraints on neutron densities, temperatures, and timescales during s-process nucleosynthesis in the parent AGB stars.

Recent measurements of the $^{96}$Mo neutron capture cross section have been performed at the n\_TOF facility (CERN) and complementary transmission measurements at GELINA (JRC-Geel) \cite{Mucciola22,Mucciola23}. The measurements utilized isotopically enriched $^{96}$Mo samples with 95.90\% enrichment. Capture cross sections were measured at both the EAR2 (18~m from spallation target) and EAR1 (184~m) experimental stations of n\_TOF during the 2021--2022 campaigns, using arrays of C$_6$D$_6$ liquid scintillation detectors and newly developed sTED detectors \cite{Alcayne24,Balibrea25}. Complementary transmission measurements were performed at the 10~m station of GELINA using a $^6$Li-glass neutron detector. 

Preliminary results show improved accuracy and consistency compared to deviations existing nuclear data libraries (JENDL-3, ENDF/B, JEFF), particularly for resonance parameters up to 5~keV. A simultaneous resonance shape analysis combining the capture and transmission data is in progress to provide final recommended resonance parameters for $^{96}$Mo. These measurements are expected to significantly reduce the current 14\% MACS uncertainty.

For $^{97}$Mo, no experimental neutron-capture cross section measurement exists to date and an experiment proposal has been approved for a high-resolution high-accuracy measurement at CERN n\_TOF \cite{Busso2024}. Given the similar nuclear structure and expected cross section magnitude compared to $^{96}$Mo, a measurement achieving comparable precision (reducing the current 15\% MACS uncertainty \cite{kadonisv1} to $\sim$4--5\%) is feasible. This would substantially improve the determination of the r-process residual, currently uncertain by $\sim$27\%.

\subsection*{$^{99}$Ru}
$^{99}$Tc, with its 2.14$\times$10$^4$~y half-life, occupies a unique position in s-process nucleosynthesis as a critical branching-point nucleus in AGB stars (see Fig.\ref{fig:MoRu}). The first neutron capture cross-section measurements by Macklin \cite{Macklin82} provided essential nuclear data for modeling the competition between neutron capture and $\beta$-decay to $^{99}$Ru during the s-process. This branching determines the flow of material through the mass region around A$\sim$100 and directly affects the final abundance distribution of ruthenium and heavier elements. The branching at $^{99}$Tc is particularly sensitive to the neutron density and temperature conditions in the He-burning shell, making the $^{99}$Ru/$^{100}$Ru abundance ratio a valuable diagnostic tool for constraining s-process conditions in AGB models.

Despite its importance as the decay product of this crucial branching point, $^{99}$Ru represents a significant gap in our nuclear data. Although the relative uncertainty quoted for the MACS of this nucleus in KADoNiS is only 15\%, the recommended neutron-capture cross section is purely based on theoretical estimates. Given that theoretical predictions more typically carry uncertainties as high as a factor of 2 even for nuclei close to stability, the quoted 15\% uncertainty is almost certainly underestimated.
Moreover, the total uncertainty in the s-process production of $^{99}$Ru is not determined solely by the $^{99}$Tc($n,\gamma$) cross section but also by the stellar $\beta$-decay rate of $^{99}$Tc. Uncertainties in stellar decay rates at this branching point can reach factors of 5-10 due to thermal population of excited states and temperature dependency \cite{Schatz83,Cosner81,Takahashi87}. For $^{99}$Tc, the competition between neutron capture and $\beta$-decay determines the branching ratio and thus the final $^{99}$Ru abundance. Both nuclear physics inputs contribute to the total uncertainty budget. Therefore, while a precision measurement of the $^{99}$Ru($n,\gamma$) cross section would reduce one component of the uncertainty, a complete understanding of the $^{99}$Ru production in AGB stars requires both, improved accuracy on the $^{99}$Ru($n,\gamma$) cross section, and better constraints on the $^{99}$Tc stellar $\beta$-decay rate. The current $\sim$32\% uncertainty on the r-process residual of $^{99}$Ru likely underestimates the true uncertainty when the branching point effect is fully considered. Nevertheless, measuring the $^{99}$Ru($n,\gamma$) cross section remains a high-priority experiment as it addresses a fundamental gap in our nuclear data-set (this is one of the few stable isotopes with no direct measurement) and provides essential input for the comprehensive s-process network calculations.

With a natural abundance of 12.76\% and a large expected cross section of 631 mb at $kT=30$ keV, producing an enriched sample of $^{99}$Ru for a time-of-flight measurement should be straightforward. Considering only the cross-section uncertainty, by reducing it to the 4-5\% level should lead to an r-process residual accurate to 6\%.

From the viewpoint of the s-process, this measurement would be extremely valuable for a fully consistent interpretation of the first detection of extinct $^{99}$Tc in presolar silicon carbide grains, which was achieved by measuring an anomaly in its decay product $^{99}$Ru using resonance ionization mass spectrometry \cite{Savina2004}. The isotopic composition of ruthenium in individual SiC stardust grains bears the signature of s-process nucleosynthesis in asymptotic giant branch stars, with an enrichment in $^{99}$Ru explained by the in situ decay of $^{99}$Tc that was present when the grains formed. This finding directly connects to Merrill's 1952 observation of technetium spectral lines in S-type stars \cite{Merrill52}, which provided the first evidence that elements are synthesized within stars. By detecting the radiogenic $^{99}$Ru signature left by now-extinct $^{99}$Tc, Savina et al. \cite{Savina2004} confirmed that the majority of presolar SiC grains originated from low-mass AGB stars and established that the amount of $^{99}$Tc produced in these stars is insufficient to leave a detectable $^{99}$Ru anomaly in bulk early solar system materials. This work allowed stellar nucleosynthesis models to be quantitatively tested with unprecedented accuracy by bridging spectroscopic observations of active stellar nucleosynthesis with laboratory measurements of preserved stellar ejecta.

\subsection*{$^{122}$Sn}
\begin{figure}[!htbp]
  \centering
  \includegraphics[width=0.8\columnwidth]{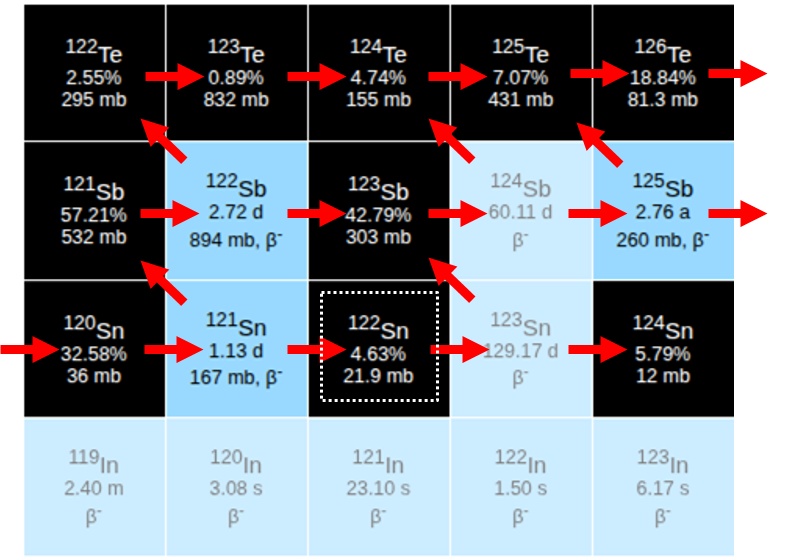}
  \caption{Tin isotopes spanning the transition from s-process-only $^{116}$Sn (not shown) to mixed r/s-process nuclei $^{122}$Sn and $^{124}$Sn.}
\label{fig:Sn}
\end{figure}

The pioneering work of Macklin and co-workers in 1962 \cite{Macklin62} demonstrated that tin's ten stable isotopes provide exceptional opportunities to disentangle the contributions of different neutron-capture processes. According to the original hypothesis of heavy-element nucleosynthesis \cite{BBFH}, certain tin isotopes can only be produced through specific pathways: $^{116}$Sn cannot form via r-process, while $^{122}$Sn and $^{124}$Sn were thought to be exclusive products of the rapid neutron capture process (see Fig.\ref{fig:Sn}). By measuring neutron capture cross-sections of the tin isotopes at stellar temperatures and examining the inverse relationship between these cross-sections and isotopic abundances it was possible to quantify for the first time the relative contributions of s-process versus r-process nucleosynthesis in the solar system's material \cite{Macklin62}. This approach confirmed theoretical predictions about how these processes operate in different stellar environments: the s-process occurring in giant stars and the r-process in explosive environments. The approximate constancy of the product of neutron capture cross-section and abundance for s-process isotopes provided direct experimental validation of nucleosynthesis models, establishing tin as an ideal ``laboratory'' element for testing our understanding of how heavy elements form in stars.

Today, thanks to detailed AGB models we know that $^{122}$Sn is not an exclusive product of the r-process \cite{Bisterzo14}. However, despite the critical role of $^{122}$Sn in disentangling s- and r-process contributions in this mass region, experimental constraints on its neutron capture cross section remain limited. The current nuclear data situation relies primarily on a single time-of-flight measurement by \cite{Timokhov1989} and two activation measurements \cite{Stadler1998,Murty1973}, one of them unpublished. The recommended MACS carries an uncertainty of 35\%, which propagates to an r-process residual uncertainty exceeding 50\% when performing classical s/r-process decomposition. This large uncertainty significantly hampers our ability to use $^{122}$Sn as a diagnostic tool for understanding the branching-point physics and for constraining galactic chemical evolution models \cite{Bisterzo14}.

Given its relatively high natural abundance of 4.63\%, $^{122}$Sn represents an excellent candidate for high-precision measurements combining time-of-flight and activation techniques. New measurements at the CERN n\_TOF facility, utilizing both the EAR1 and EAR2 beam lines as well as the NEAR station for activation studies \cite{Patronis25}, would provide the complementary data needed to substantially reduce these uncertainties and better constrain the s-process contribution to this key branching-point isotope.

\subsection*{$^{139}$La}
\begin{figure}[!htbp]
  \centering
  \includegraphics[width=0.8\columnwidth]{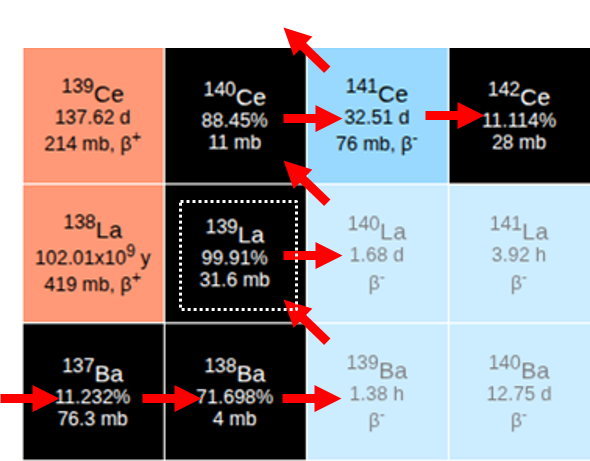}
  \caption{Lanthanum isotopes near the N = 82 magic number, where $^{139}$La acts as an s-process bottleneck.}
\label{fig:La}
\end{figure}

$^{139}$La is the most abundant isotope of lanthanum ($>99.9$\%) and is primarily produced in the s-process, with an estimated contribution of 75\% \cite{Bisterzo14}. It is of particular interest since it has a magic neutron number ($N=82$) and as a consequence a small neutron capture cross section compared to neighbouring isotopes (see Fig.\ref{fig:La}). Hence, the $^{139}$La($n,\gamma$) reaction represents a bottleneck in the s-process flow where matter accumulates. Based on the current uncertainties of the Maxwellian averaged cross section of around 10\%, the resultant uncertainty of the solar r-process abundance is around 39\%. \\

There are several measurements of the $^{139}$La($n,\gamma$) cross section \cite{kadonisv1}. While available experimental MACS values at $kT=30$~keV range from 22.2 to 50 mb \cite{Dillmann08}, the three most recent results quote consistent values of $31.6\pm0.8$~mb \cite{Obrien03}, $31.2$ mb (renormalised from a measurement at $kT=5$~keV) \cite{Winckler_2006} and $32.4\pm2.1$~mb \cite{Terlizzi07}. The recommended value in the KADoNiS database and hence the value most frequently used in stellar model calculations is the result from measurement at n\_TOF by Terlizzi et al \cite{Terlizzi07}. This measurement covered neutron energy ranges from  0.6 eV to 9 keV. The Maxwellian averaged cross sections were calculated by using the extracted resonance information combined with the evaluated cross sections from the JENDL-3.3 database between 9~keV and 1~MeV, which has been scaled to match the experimental cross section at lower neutron energy. The total uncertainty of the MACS at 30~keV is dominated by the uncertainty assumed on the JENDL-3.3 cross section (10\%). The activation measurement by O'Brien et al. \cite{Obrien03} is the only direct measurement at 30~keV neutron energy. The quoted uncertainty is small and does not include the uncertainty of the $^{197}$Au($n,\gamma$) the spectrum averaged cross section used as a reference reaction. Incidentally, the $^{197}$Au($n,\gamma$) in the keV neutron energy region has recently been updated \cite{Lederer11,Massimi14,Reifarth2018}, which would lead to a small reduction of the value obtained by \cite{Obrien03}. \\
Concerning the n\_TOF measurement \cite{Terlizzi07}, the energy range analysed was limited by the increase of in-beam $\gamma$-ray background in the keV neutron energy range,  combined with low statistics due to the small neutron capture cross section. Since then, the spallation target at n\_TOF was replaced, and now borated water is used as neutron moderator. This results in a significant reduction of the in-beam $\gamma$-ray background (see Fig.13 in Ref.\cite{Massimi10}, and Refs. \cite{Lerendegui16a,Lerendegui16b}), and neutron capture cross sections have since been routinely measured up to tens or hundreds of keV, even for comparably small cross sections, e.g.  \cite{Lederer14,Gawlik21,Amaducci24}. A new measurement, aiming for high counting statistics at n\_TOF would allow to access the entire neutron energy range relevant for AGB stars (the main producers of the s-process component of $^{139}$La), and reduce uncertainties to 5-6\%, typically achieved at n\_TOF. This can be combined with an activation measurement at n\_TOF NEAR. In parallel, a re-normalisation of the O'Brien \cite{Obrien03} values based on the updated $^{197}$Au($n,\gamma$) cross section will help to provide a second independent value for the MACS at $kT=30$~keV. 

\subsection*{$^{168}$Er}
\begin{figure}[!htbp]
  \centering
  \includegraphics[width=0.8\columnwidth]{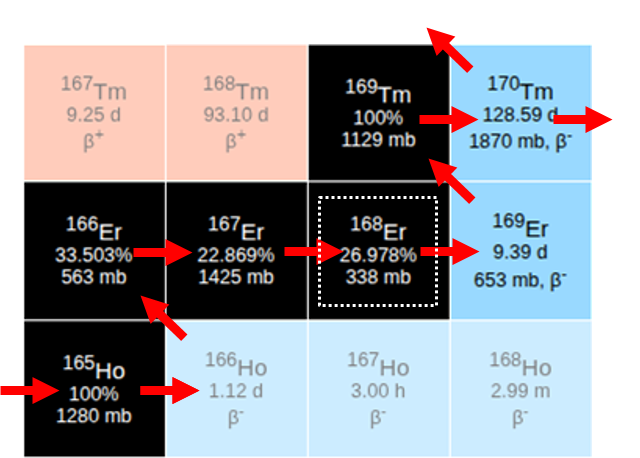}
  \caption{Erbium isotopes in the deformed mid-shell region between the second and third s-process abundance peaks.}
\label{fig:Er}
\end{figure}

Erbium isotopes, including $^{168}$Er, occupy a critical region in the landscape of s-process nucleosynthesis (see Fig.\ref{fig:Er}). Located beyond the N = 82 neutron magic number in the mass range A $\sim$ 166--170, the erbium isotopes lie in the transition zone between the second s-process abundance peak (A $\sim$ 140) and the third peak (A $\sim$ 208). 

The s-process dominates the production of elements in the erbium-ytterbium (Er--Yb) region, with contributions of 70--90\% of the solar system abundances. This mass region is characterized by deformed, mid-shell nuclei with low-lying excited states that can significantly impact stellar neutron capture rates through thermal population effects. The relatively large uncertainties in the MACS for  $^{168}$Er represents a limiting factor in the precision ($\sim$18\%) of s-process low-mass AGB models, which propagates to almost 30\% for the corresponding r-process component. With $^{168}$Er having a relatively large cross section (>300 mb at 30keV) and a natural abundance of approximately 27\%, a new TOF measurement at n\_TOF EAR1 or EAR2 should be able to deliver new MACS values with an uncertainty down to $\sim$5\% in the 30 keV energy range. 

\subsection*{$^{184}$W}
\begin{figure}[!htbp]
  \centering
  \includegraphics[width=0.8\columnwidth]{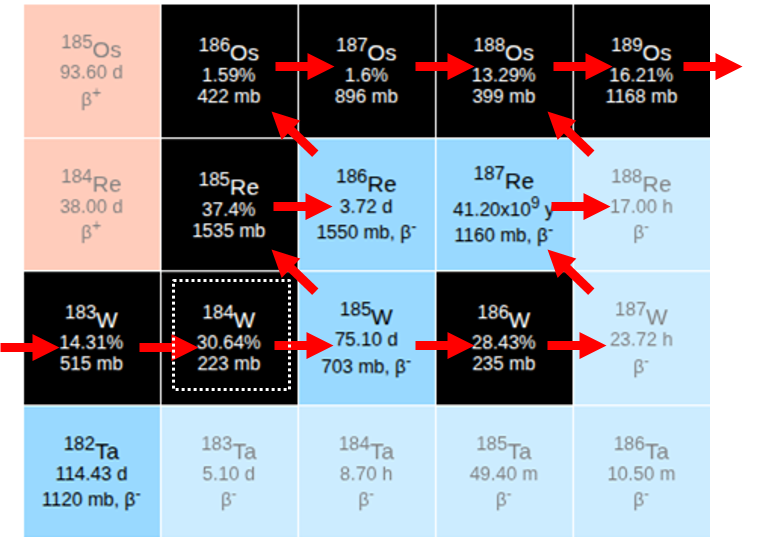}
  \caption{Tungsten isotopes near the $^{185}$W branching point, critical for Re/Os cosmochronometry.}
\label{fig:W}
\end{figure}

The neutron-capture cross section of $^{184}$W plays a crucial role in s-process nucleosynthesis, determining the abundance flow through the tungsten isotope chain in AGB stars (see Fig.\ref{fig:W}). While the branching at $^{185}$W is directly important for the Re/Os cosmochronometer \cite{Mosconi2010}, accurate knowledge of the $^{184}$W(n,$\gamma$) cross section is essential for modeling the s-process reaction network through this mass region and for understanding $^{182}$Hf/$^{182}$W chronometry of the early solar system. As the second-most neutron-rich stable tungsten isotope, $^{184}$W is particularly valuable for constraining the cross-section trend with neutron number, which is critical for extrapolating statistical model calculations to unstable nuclei relevant for explosive nucleosynthesis (r- and p-processes).

The most recent direct measurement \cite{Marganiec09} employed the activation technique at the Karlsruhe Van de Graaff accelerator, yielding a MACS at $kT=30$ keV of $221 \pm 12$ mb (5.4\% uncertainty). This value confirmed the previously recommended MACS of $224 \pm 10$ mb \cite{Bao00} and showed good agreement with earlier time-of-flight measurements \cite{Macklin83,Beer82}. However, when combined with new data for the neighboring isotopes $^{180}$W and $^{186}$W, the measurements revealed a significantly flatter cross-section trend with mass number than predicted by Hauser-Feshbach calculations \cite{Rauscher00,Goriely}, highlighting remaining challenges in theoretical modeling and indicating that statistical model predictions commonly used for r-process network calculations may overestimate cross sections for the neutron-rich tungsten isotopes.

From the viewpoint of enhanced accuracy measurements, improving the presently recommended relative uncertainty of $\sim$12\%\cite{kadonisv1} down to 4\% could still help to reduce the 57\% uncertainty on the r-process isotopic abundance down to $\sim$20\%. Such a measurement at CERN n\_TOF could benefit of the availability of both TOF- and activation techniques and new instrumentation at EAR1, EAR2 \cite{Alcayne24,Balibrea25} and NEAR \cite{Patronis25} stations. A measurement which, considering the relatively large cross section and natural abundance of $^{184}$W (30\%) should be rather straight-forward.

\subsection*{$^{200}$Hg}

\begin{figure}[!htbp]
  \centering
  \includegraphics[width=0.9\columnwidth]{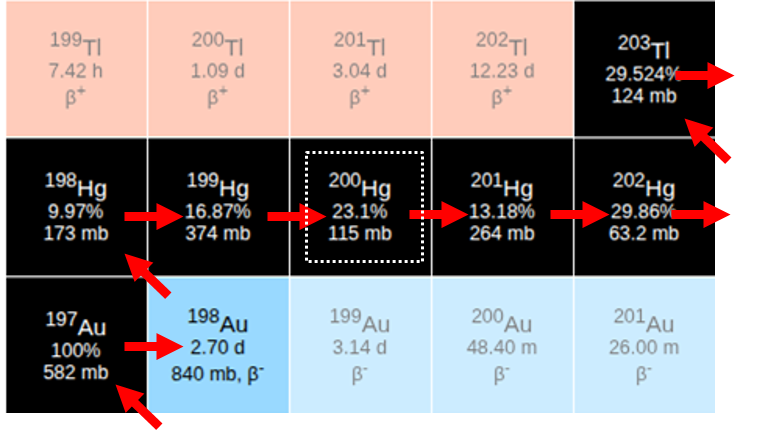}
  \caption{Mercury isotopes in the A$\sim$200 plateau region, showing balanced s- and r-process contributions.}
\label{fig:Hg}
\end{figure}

The r-process residual derived from $^{200}$Hg may serve as an important anchor point for constraining the r-process abundance distribution in the $A \sim 200$ mass region (see Fig.\ref{fig:Hg}). Together with the other mercury isotopes ($^{199,201,202}$Hg), the $^{200}$Hg data help to define the relatively flat r-process abundance plateau expected in this mass range, lying between the $N=126$ r-process peak and the lead region. This information is helpful for validating r-process nucleosynthesis models and for understanding the conditions under which rapid neutron capture occurred in the early Galaxy.

With a measured MACS of $115 \pm 12$ mb at $kT=30$ keV, $^{200}$Hg represents a key (s+r) process isotope that receives rather balanced contributions from both slow and rapid neutron-capture processes. Thus, the accurate cross section determination enables quantitative decomposition of the $^{200}$Hg solar abundance into its s- and r-process components. Using the measured cross section and the $\sigma N$ systematics of the s-process, Beer and Macklin determined that of the total solar abundance of $(0.045 \pm 0.007)/10^6$ Si, the s-process contributes 42(5)\% while the r-process accounts for 55(16)\%\cite{Beer85}. A new measurements could significantly decrease the uncertainty, down to $\sim$5\%, thus lowering also the relative r-process uncertainty down to the level of $\sim$10\%.

\section{Summary and conclusions}

We have presented a systematic analysis of neutron capture cross section uncertainties and their impact on r-process residual determinations for isotopes in the mass range $A \sim 96$--200. By combining MACS data from KADoNiS 1.0 \cite{kadonisv1}, solar system abundances from \cite{Lodders2025}, and s-process predictions from \cite{Bisterzo14}, we have identified a set of high-priority targets where improved measurements would significantly reduce uncertainties in derived r-process yields.

Our analysis reveals that seven isotopes meet the dual criteria of MACS relative uncertainty $\geq 10\%$ and r-process residual relative uncertainty $\geq 20\%$: $^{97}$Mo, $^{99}$Ru, $^{122}$Sn, $^{139}$La, $^{168}$Er, $^{184}$W and $^{200}$Hg. These isotopes span critical regions of the chart of nuclides, including the $N = 50$ shell closure ($^{97}$Mo), s-process branching points at $A = 122$--123 ($^{122}$Sn), the neutron magic number $N = 82$ ($^{139}$La), and the rare earth region where main s-process contributions dominate ($^{168}$Er). The case of $^{99}$Ru is particularly noteworthy, as current KADoNiS recommendations rely entirely on theoretical calculations with no direct experimental constraints.

The enhanced capabilities of CERN n\_TOF following Long Shutdown 3 present an excellent opportunity to address these measurement needs. The combination of the high-flux EAR1 and EAR2 \cite{Pavon25} beam lines with the newly commissioned NEAR station \cite{Patronis25} provides optimal conditions for both time-of-flight and activation measurements across the relevant energy range. The improved neutron flux, enhanced detection systems \cite{Alcayne24,Balibrea25}, and extended measurement capabilities will enable precision determinations of MACS values with target uncertainties below 5\% for most of the identified isotopes.

Beyond the immediate goal of constraining r-process residuals, these measurements will provide critical input for s-process nucleosynthesis models themselves. Isotopes such as $^{122}$Sn and $^{139}$La are located at key branching points and bottlenecks in the s-process path, where neutron capture rates directly influence the final abundance distributions produced in AGB stars. Improved cross sections for these nuclei will therefore refine our understanding of both the weak and main s-process components, with implications for galactic chemical evolution models and the interpretation of abundance patterns in metal-poor stars.

The linear uncertainty propagation approach employed in this work, while simplified, provides a practical framework for prioritizing experimental efforts. Future refinements could incorporate full network calculations to account for correlation effects and non-linear sensitivities, particularly for branching point isotopes where competition between neutron capture and $\beta$-decay determines the s-process flow. Nevertheless, our first-order analysis successfully identifies the most critical cases where nuclear data improvements will have maximum impact.

In conclusion, the post-LS3 experimental program at CERN n\_TOF represents a significant step forward in addressing long-standing uncertainties in neutron capture cross sections that limit our ability to cleanly separate s-process and r-process contributions to solar system abundances. The measurements proposed here will not only improve constraints on r-process nucleosynthesis yields but will also advance our understanding of s-process physics in AGB stars, contributing to a more complete picture of heavy element production in the Galaxy.

\section*{Acknowledgments}
We dedicate this work to the memory of Roberto Gallino, whose pioneering 
theoretical work on s-process nucleosynthesis in AGB stars provided the 
essential framework for interpreting neutron-capture measurements such as 
those performed at FZK Karlsruhe and CERN n\_TOF \cite{Gallino23}. His detailed stellar models, incorporating 
the physics of $^{13}$C-pockets and nuclear-physics networks, have been 
instrumental in calculating the s-process contributions needed to understand the composition of our solar system and the chemical evolution of our galaxy. His legacy continues to guide our efforts to constrain the origins of the heavy elements in the cosmos.

CDP acknowledges fruitful discussions with Almudena Arcones and Gabriel Martínez Pinedo at the EMMI Workshop "Nucleosynthesis of heavy elements: r-process" held in Hirschegg, Austria, in 2025. The authors acknowledge K. Lodders and I. Dillmann for facilitating quick access to the data of their publications and evaluations. Fig.\,\ref{fig:MoRu} to Fig.\,\ref{fig:Hg} have been adapted from the KADoNiS web database.

The authors acknowledge support from all the funding agencies of participating institutions. Part of this work was supported by the MCIN/AEI 10.13039/\-501100011033 under grants Severo Ochoa CEX\-2023-001\-292\--S and PID2022-138297NB-C21 and UKRI STFC Grant ST/Y000293/1.
The corresponding author has applied a Creative Commons Attribution (CC BY-NC-ND 4.0) license.

\section*{Supplemental material}

\begin{table*}[t]
\centering
\caption{Solar system isotopic abundances ($N_{\odot}$), s-process contributions ($N_s$), and derived r-process residuals ($N_r = N_{\odot} - N_s$) for stable isotopes in the mass range A = 80–209. Abundances are normalized to Si$ = 10^6$. Solar abundances are from Lodders et al.\cite{Lodders2025}, s-process contributions are from the galactic chemical evolution models of Bisterzo et al. \cite{Bisterzo14}. The columns $N_r/N_\odot$ and $N_s/N_\odot$ represent the percentage contributions of r-process and s-process to the total solar abundance, respectively. Uncertainties on r-process residuals are derived from the propagation of MACS uncertainties ($\Delta \sigma/\langle \sigma \rangle$, last column) using the linear approximation described in Section \ref{sec:methodology}, Equation (3). Values in parentheses represent uncertainties in the last significant digits.}
\label{tab:isotope_abundances}
\begin{tabular}{lcccccc}
\hline\hline
Isotope & $N_{\odot}$ & $N_r$ & $N_r/N_{\odot}$ & $N_s$ & $N_s/N_{\odot}$ & $\Delta\sigma/\langle\sigma\rangle$ \\
 & $(\times 10^6~\mathrm{Si})$ & $(\times 10^6~\mathrm{Si})$ & (\%) & $(\times 10^6~\mathrm{Si})$ & (\%) & (\%) \\
\hline
$^{80}$Kr & $1.34$ & $1.16(7)$ & 87 & $1.78(73) \times 10^{-1}$ & 13 & 5 \\
$^{82}$Kr & $6.70$ & $4.86(45)$ & 73 & $1.84(45)$ & 27 & 7 \\
$^{83}$Kr & $6.63$ & $6.00(35)$ & 90 & $6.30(346) \times 10^{-1}$ & 10 & 5 \\
$^{84}$Kr & $3.26 \times 10^{1}$ & $2.88(25) \times 10^{1}$ & 88 & $3.81(246)$ & 12 & 8 \\
$^{86}$Kr & $9.82$ & $8.33(58)$ & 85 & $1.49(58)$ & 15 & 6 \\
$^{85}$Rb & $5.02$ & $4.25(19)$ & 85 & $7.68(188) \times 10^{-1}$ & 15 & 4 \\
$^{87}$Rb & $2.07$ & $1.55(11)$ & 75 & $5.15(110) \times 10^{-1}$ & 25 & 5 \\
$^{86}$Sr & $2.37$ & $9.22(178) \times 10^{-1}$ & 39 & $1.45(18)$ & 61 & 8 \\
$^{87}$Sr & $1.66$ & $6.94(67) \times 10^{-1}$ & 42 & $9.66(67) \times 10^{-1}$ & 58 & 4 \\
$^{88}$Sr & $1.98 \times 10^{1}$ & $5.72(63)$ & 29 & $1.41(6) \times 10^{1}$ & 71 & 3 \\
$^{89}$Y & $4.49$ & $1.26(13)$ & 28 & $3.23(13)$ & 72 & 3 \\
$^{90}$Zr & $5.41$ & $2.15(25)$ & 40 & $3.26(25)$ & 60 & 5 \\
$^{91}$Zr & $1.18$ & $3.40(75) \times 10^{-1}$ & 29 & $8.40(75) \times 10^{-1}$ & 71 & 6 \\
$^{92}$Zr & $1.80$ & $5.72(143) \times 10^{-1}$ & 32 & $1.23(14)$ & 68 & 8 \\
$^{94}$Zr & $1.82$ & $2.98(69) \times 10^{-1}$ & 16 & $1.52(7)$ & 84 & 4 \\
$^{96}$Zr & $2.90 \times 10^{-1}$ & $1.78(18) \times 10^{-1}$ & 61 & $1.12(18) \times 10^{-1}$ & 39 & 6 \\
$^{93}$Nb & $7.66 \times 10^{-1}$ & $3.37(15) \times 10^{-1}$ & 44 & $4.29(15) \times 10^{-1}$ & 56 & 2 \\
$^{94}$Mo & $2.38 \times 10^{-1}$ & $2.36(32) \times 10^{-1}$ & 99 & $2.38(3216) \times 10^{-3}$ & 1 & 14 \\
$^{95}$Mo & $4.12 \times 10^{-1}$ & $2.25(43) \times 10^{-1}$ & 55 & $1.87(43) \times 10^{-1}$ & 45 & 10 \\
$^{96}$Mo & $4.33 \times 10^{-1}$ & $9.44(619) \times 10^{-2}$ & 22 & $3.39(62) \times 10^{-1}$ & 78 & 14 \\
$^{97}$Mo & $2.49 \times 10^{-1}$ & $1.41(38) \times 10^{-1}$ & 57 & $1.08(38) \times 10^{-1}$ & 43 & 15 \\
$^{98}$Mo & $6.30 \times 10^{-1}$ & $2.68(42) \times 10^{-1}$ & 43 & $3.62(42) \times 10^{-1}$ & 58 & 7 \\
$^{100}$Mo & $2.53 \times 10^{-1}$ & $2.47(18) \times 10^{-1}$ & 98 & $5.82(1788) \times 10^{-3}$ & 2 & 7 \\
$^{99}$Ru & $2.26 \times 10^{-1}$ & $1.79(57) \times 10^{-1}$ & 79 & $4.75(566) \times 10^{-2}$ & 21 & 25 \\
$^{100}$Ru & $2.23 \times 10^{-1}$ & $4.44(147) \times 10^{-2}$ & 20 & $1.79(15) \times 10^{-1}$ & 80 & 7 \\
$^{101}$Ru & $3.02 \times 10^{-1}$ & $2.63(20) \times 10^{-1}$ & 87 & $3.87(199) \times 10^{-2}$ & 13 & 7 \\
$^{102}$Ru & $5.58 \times 10^{-1}$ & $3.16(36) \times 10^{-1}$ & 57 & $2.42(36) \times 10^{-1}$ & 43 & 6 \\
$^{104}$Ru & $3.29 \times 10^{-1}$ & $3.24(16) \times 10^{-1}$ & 99 & $4.61(1625) \times 10^{-3}$ & 1 & 5 \\
$^{103}$Rh & $3.41 \times 10^{-1}$ & $3.01(6) \times 10^{-1}$ & 88 & $4.02(57) \times 10^{-2}$ & 12 & 2 \\
$^{104}$Pd & $1.54 \times 10^{-1}$ & $2.48(33) \times 10^{-2}$ & 16 & $1.29(3) \times 10^{-1}$ & 84 & 2 \\
$^{105}$Pd & $3.09 \times 10^{-1}$ & $2.76(5) \times 10^{-1}$ & 89 & $3.31(54) \times 10^{-2}$ & 11 & 2 \\
$^{106}$Pd & $3.79 \times 10^{-1}$ & $2.28(38) \times 10^{-1}$ & 60 & $1.51(38) \times 10^{-1}$ & 40 & 10 \\
$^{108}$Pd & $3.67 \times 10^{-1}$ & $1.82(28) \times 10^{-1}$ & 50 & $1.85(28) \times 10^{-1}$ & 50 & 8 \\
$^{110}$Pd & $1.62 \times 10^{-1}$ & $1.59(26) \times 10^{-1}$ & 98 & $2.59(2570) \times 10^{-3}$ & 2 & 16 \\
$^{107}$Ag & $2.61 \times 10^{-1}$ & $2.61(8) \times 10^{-1}$ & 100 & $2.61(7879) \times 10^{-4}$ & 0 & 3 \\
$^{109}$Ag & $2.42 \times 10^{-1}$ & $1.87(7) \times 10^{-1}$ & 77 & $5.47(71) \times 10^{-2}$ & 23 & 3 \\
$^{108}$Cd & $1.40 \times 10^{-2}$ & $1.38(6) \times 10^{-2}$ & 99 & $1.54(642) \times 10^{-4}$ & 1 & 5 \\
$^{110}$Cd & $1.99 \times 10^{-1}$ & $4.46(35) \times 10^{-2}$ & 22 & $1.54(3) \times 10^{-1}$ & 78 & 2 \\
$^{111}$Cd & $2.04 \times 10^{-1}$ & $1.53(3) \times 10^{-1}$ & 75 & $5.08(34) \times 10^{-2}$ & 25 & 2 \\
$^{112}$Cd & $3.85 \times 10^{-1}$ & $1.93(3) \times 10^{-1}$ & 50 & $1.92(3) \times 10^{-1}$ & 50 & 1 \\
$^{113}$Cd & $1.95 \times 10^{-1}$ & $1.39(10) \times 10^{-1}$ & 72 & $5.56(98) \times 10^{-2}$ & 28 & 5 \\
$^{114}$Cd & $4.58 \times 10^{-1}$ & $1.87(5) \times 10^{-1}$ & 41 & $2.71(5) \times 10^{-1}$ & 59 & 1 \\
$^{116}$Cd & $1.20 \times 10^{-1}$ & $1.09(1) \times 10^{-1}$ & 91 & $1.06(14) \times 10^{-2}$ & 9 & 1 \\
$^{113}$In & $8.00 \times 10^{-3}$ & $8.00(112) \times 10^{-3}$ & 100 & $0$ & 0 & 14 \\
$^{115}$In & $1.71 \times 10^{-1}$ & $1.20(18) \times 10^{-1}$ & 70 & $5.10(181) \times 10^{-2}$ & 30 & 11 \\
$^{114}$Sn & $2.40 \times 10^{-2}$ & $2.40(3) \times 10^{-2}$ & 100 & $2.40(3214) \times 10^{-5}$ & 0 & 1 \\
$^{115}$Sn & $1.30 \times 10^{-2}$ & $1.25(3) \times 10^{-2}$ & 96 & $5.46(331) \times 10^{-4}$ & 4 & 3 \\
$^{116}$Sn & $5.37 \times 10^{-1}$ & $1.70(12) \times 10^{-1}$ & 32 & $3.67(12) \times 10^{-1}$ & 68 & 2 \\
$^{117}$Sn & $2.83 \times 10^{-1}$ & $1.74(16) \times 10^{-1}$ & 62 & $1.09(16) \times 10^{-1}$ & 38 & 6 \\
$^{118}$Sn & $8.94 \times 10^{-1}$ & $3.99(9) \times 10^{-1}$ & 45 & $4.95(9) \times 10^{-1}$ & 55 & 1 \\
$^{119}$Sn & $3.17 \times 10^{-1}$ & $1.68(13) \times 10^{-1}$ & 53 & $1.49(13) \times 10^{-1}$ & 47 & 4 \\
$^{120}$Sn & $1.20$ & $4.49(30) \times 10^{-1}$ & 37 & $7.54(30) \times 10^{-1}$ & 63 & 2 \\
$^{122}$Sn & $1.71 \times 10^{-1}$ & $1.09(59) \times 10^{-1}$ & 64 & $6.22(590) \times 10^{-2}$ & 36 & 35 \\
$^{121}$Sb & $1.94 \times 10^{-1}$ & $1.34(6) \times 10^{-1}$ & 69 & $5.99(58) \times 10^{-2}$ & 31 & 3 \\
$^{123}$Sb & $1.45 \times 10^{-1}$ & $1.38(4) \times 10^{-1}$ & 95 & $7.25(431) \times 10^{-3}$ & 5 & 3 \\

\hline
\end{tabular}
\end{table*}

\newpage

\begin{table*}[t]
\centering
\caption{[Continuation]}
\label{tab:isotope_abundances}
\begin{tabular}{lcccccc}
\hline\hline
Isotope & $N_{\odot}$ & $N_r$ & $N_r/N_{\odot}$ & $N_s$ & $N_s/N_{\odot}$ & $\Delta\sigma/\langle\sigma\rangle$ \\
 & $(\times 10^6~\mathrm{Si})$ & $(\times 10^6~\mathrm{Si})$ & (\%) & $(\times 10^6~\mathrm{Si})$ & (\%) & (\%) \\
\hline
$^{122}$Te & $1.24 \times 10^{-1}$ & $3.27(11) \times 10^{-2}$ & 26 & $9.13(11) \times 10^{-2}$ & 74 & 1 \\
$^{123}$Te & $4.30 \times 10^{-2}$ & $1.11(4) \times 10^{-2}$ & 26 & $3.19(4) \times 10^{-2}$ & 74 & 1 \\
$^{124}$Te & $2.29 \times 10^{-1}$ & $5.02(30) \times 10^{-2}$ & 22 & $1.79(3) \times 10^{-1}$ & 78 & 1 \\
$^{125}$Te & $3.39 \times 10^{-1}$ & $2.80(3) \times 10^{-1}$ & 83 & $5.90(31) \times 10^{-2}$ & 17 & 1 \\
$^{126}$Te & $8.99 \times 10^{-1}$ & $5.73(15) \times 10^{-1}$ & 64 & $3.26(15) \times 10^{-1}$ & 36 & 2 \\
$^{128}$Te & $1.50$ & $1.45(4)$ & 97 & $4.96(440) \times 10^{-2}$ & 3 & 3 \\
$^{127}$I & $1.59$ & $1.53(21)$ & 96 & $6.04(2138) \times 10^{-2}$ & 4 & 13 \\
$^{128}$Xe & $1.24 \times 10^{-1}$ & $2.94(17) \times 10^{-2}$ & 24 & $9.46(17) \times 10^{-2}$ & 76 & 1 \\
$^{129}$Xe & $1.54$ & $1.49(3)$ & 97 & $4.30(299) \times 10^{-2}$ & 3 & 2 \\
$^{130}$Xe & $2.44 \times 10^{-1}$ & $3.64(39) \times 10^{-2}$ & 15 & $2.08(4) \times 10^{-1}$ & 85 & 2 \\
$^{132}$Xe & $1.48$ & $1.08(12)$ & 73 & $3.96(123) \times 10^{-1}$ & 27 & 8 \\
$^{134}$Xe & $5.45 \times 10^{-1}$ & $5.23(49) \times 10^{-1}$ & 96 & $2.23(486) \times 10^{-2}$ & 4 & 9 \\
$^{133}$Cs & $3.67 \times 10^{-1}$ & $3.17(20) \times 10^{-1}$ & 86 & $4.95(205) \times 10^{-2}$ & 14 & 6 \\
$^{134}$Ba & $1.12 \times 10^{-1}$ & $1.11(4) \times 10^{-1}$ & 99 & $1.01(356) \times 10^{-3}$ & 1 & 3 \\
$^{135}$Ba & $3.04 \times 10^{-1}$ & $2.17(10) \times 10^{-1}$ & 72 & $8.66(100) \times 10^{-2}$ & 28 & 3 \\
$^{136}$Ba & $3.62 \times 10^{-1}$ & $3.57(12) \times 10^{-1}$ & 99 & $4.71(1185) \times 10^{-3}$ & 1 & 3 \\
$^{137}$Ba & $5.18 \times 10^{-1}$ & $1.91(17) \times 10^{-1}$ & 37 & $3.27(17) \times 10^{-1}$ & 63 & 3 \\
$^{138}$Ba & $3.31$ & $2.71(160) \times 10^{-1}$ & 8 & $3.04(16)$ & 92 & 5 \\
$^{139}$La & $4.68 \times 10^{-1}$ & $1.15(45) \times 10^{-1}$ & 24 & $3.53(45) \times 10^{-1}$ & 76 & 10 \\
$^{140}$Ce & $1.05$ & $8.40(394) \times 10^{-2}$ & 8 & $9.66(39) \times 10^{-1}$ & 92 & 4 \\
$^{142}$Ce & $1.32 \times 10^{-1}$ & $1.06(4) \times 10^{-1}$ & 80 & $2.57(44) \times 10^{-2}$ & 20 & 3 \\
$^{141}$Pr & $1.79 \times 10^{-1}$ & $8.97(22) \times 10^{-2}$ & 50 & $8.93(22) \times 10^{-2}$ & 50 & 1 \\
$^{142}$Nd & $2.37 \times 10^{-1}$ & $4.03(473) \times 10^{-3}$ & 2 & $2.33(5) \times 10^{-1}$ & 98 & 2 \\
$^{143}$Nd & $1.05 \times 10^{-1}$ & $7.05(13) \times 10^{-2}$ & 67 & $3.45(13) \times 10^{-2}$ & 33 & 1 \\
$^{144}$Nd & $2.08 \times 10^{-1}$ & $9.94(38) \times 10^{-2}$ & 48 & $1.09(4) \times 10^{-1}$ & 52 & 2 \\
$^{145}$Nd & $7.70 \times 10^{-2}$ & $5.68(8) \times 10^{-2}$ & 74 & $2.02(8) \times 10^{-2}$ & 26 & 1 \\
$^{146}$Nd & $1.50 \times 10^{-1}$ & $5.10(16) \times 10^{-2}$ & 34 & $9.90(16) \times 10^{-2}$ & 66 & 1 \\
$^{148}$Nd & $5.00 \times 10^{-2}$ & $4.13(6) \times 10^{-2}$ & 83 & $8.65(65) \times 10^{-3}$ & 17 & 1 \\
$^{148}$Sm & $4.22 \times 10^{-2}$ & $4.08(11) \times 10^{-2}$ & 97 & $1.43(114) \times 10^{-3}$ & 3 & 3 \\
$^{149}$Sm & $3.07 \times 10^{-2}$ & $2.67(7) \times 10^{-2}$ & 87 & $3.96(74) \times 10^{-3}$ & 13 & 2 \\
$^{150}$Sm & $3.77 \times 10^{-2}$ & $0$ & 0 & $3.77(4) \times 10^{-2}$ & 100 & 1 \\
$^{152}$Sm & $2.01 \times 10^{-2}$ & $1.55(2) \times 10^{-2}$ & 77 & $4.58(19) \times 10^{-3}$ & 23 & 1 \\
$^{154}$Sm & $7.29 \times 10^{-2}$ & $7.11(33) \times 10^{-2}$ & 98 & $1.82(326) \times 10^{-3}$ & 2 & 4 \\
$^{151}$Eu & $4.98 \times 10^{-2}$ & $4.69(15) \times 10^{-2}$ & 94 & $2.94(148) \times 10^{-3}$ & 6 & 3 \\
$^{153}$Eu & $5.43 \times 10^{-2}$ & $5.10(37) \times 10^{-2}$ & 94 & $3.31(374) \times 10^{-3}$ & 6 & 7 \\
$^{152}$Gd & $7.10 \times 10^{-4}$ & $9.51(115) \times 10^{-5}$ & 13 & $6.15(12) \times 10^{-4}$ & 87 & 2 \\
$^{154}$Gd & $7.68 \times 10^{-3}$ & $8.37(90) \times 10^{-4}$ & 11 & $6.84(9) \times 10^{-3}$ & 89 & 1 \\
$^{155}$Gd & $5.21 \times 10^{-2}$ & $4.90(6) \times 10^{-2}$ & 94 & $3.13(59) \times 10^{-3}$ & 6 & 1 \\
$^{156}$Gd & $7.21 \times 10^{-2}$ & $5.92(6) \times 10^{-2}$ & 82 & $1.29(6) \times 10^{-2}$ & 18 & 1 \\
$^{157}$Gd & $5.51 \times 10^{-2}$ & $4.90(6) \times 10^{-2}$ & 89 & $6.12(60) \times 10^{-3}$ & 11 & 1 \\
$^{158}$Gd & $8.75 \times 10^{-2}$ & $6.34(8) \times 10^{-2}$ & 72 & $2.41(8) \times 10^{-2}$ & 28 & 1 \\
$^{160}$Gd & $7.70 \times 10^{-2}$ & $7.65(108) \times 10^{-2}$ & 99 & $5.39(10816) \times 10^{-4}$ & 1 & 14 \\
$^{159}$Tb & $6.37 \times 10^{-2}$ & $5.86(90) \times 10^{-2}$ & 92 & $5.10(904) \times 10^{-3}$ & 8 & 14 \\
$^{160}$Dy & $9.80 \times 10^{-3}$ & $9.80(132) \times 10^{-4}$ & 10 & $8.82(13) \times 10^{-3}$ & 90 & 1 \\
$^{161}$Dy & $7.93 \times 10^{-2}$ & $7.51(8) \times 10^{-2}$ & 95 & $4.20(77) \times 10^{-3}$ & 5 & 1 \\
$^{162}$Dy & $1.07 \times 10^{-1}$ & $8.98(10) \times 10^{-2}$ & 84 & $1.71(10) \times 10^{-2}$ & 16 & 1 \\
$^{163}$Dy & $1.05 \times 10^{-1}$ & $9.99(10) \times 10^{-2}$ & 96 & $4.60(103) \times 10^{-3}$ & 4 & 1 \\
$^{164}$Dy & $1.19 \times 10^{-1}$ & $9.03(17) \times 10^{-2}$ & 76 & $2.83(17) \times 10^{-2}$ & 24 & 1 \\
$^{165}$Ho & $9.06 \times 10^{-2}$ & $8.31(134) \times 10^{-2}$ & 92 & $7.52(1340) \times 10^{-3}$ & 8 & 15 \\
$^{164}$Er & $4.00 \times 10^{-3}$ & $6.64(169) \times 10^{-4}$ & 17 & $3.34(17) \times 10^{-3}$ & 83 & 4 \\
$^{166}$Er & $8.80 \times 10^{-2}$ & $7.33(94) \times 10^{-2}$ & 83 & $1.47(94) \times 10^{-2}$ & 17 & 11 \\
$^{167}$Er & $6.00 \times 10^{-2}$ & $5.44(96) \times 10^{-2}$ & 91 & $5.64(960) \times 10^{-3}$ & 9 & 16 \\
$^{168}$Er & $7.10 \times 10^{-2}$ & $4.87(129) \times 10^{-2}$ & 69 & $2.23(129) \times 10^{-2}$ & 31 & 18 \\
$^{170}$Er & $3.90 \times 10^{-2}$ & $3.41(16) \times 10^{-2}$ & 87 & $4.91(156) \times 10^{-3}$ & 13 & 4 \\
$^{169}$Tm & $4.10 \times 10^{-2}$ & $3.73(25) \times 10^{-2}$ & 91 & $3.73(250) \times 10^{-3}$ & 9 & 6 \\
$^{170}$Yb & $8.00 \times 10^{-3}$ & $3.36(75) \times 10^{-4}$ & 4 & $7.66(7) \times 10^{-3}$ & 96 & 1 \\
$^{171}$Yb & $3.60 \times 10^{-2}$ & $2.78(4) \times 10^{-2}$ & 77 & $8.17(36) \times 10^{-3}$ & 23 & 1 \\
$^{172}$Yb & $5.60 \times 10^{-2}$ & $3.14(5) \times 10^{-2}$ & 56 & $2.46(5) \times 10^{-2}$ & 44 & 1 \\
$^{173}$Yb & $4.20 \times 10^{-2}$ & $3.03(4) \times 10^{-2}$ & 72 & $1.17(4) \times 10^{-2}$ & 28 & 1 \\
$^{174}$Yb & $8.30 \times 10^{-2}$ & $3.25(10) \times 10^{-2}$ & 39 & $5.05(10) \times 10^{-2}$ & 61 & 1 \\
$^{176}$Yb & $3.40 \times 10^{-2}$ & $3.16(6) \times 10^{-2}$ & 93 & $2.45(59) \times 10^{-3}$ & 7 & 2 \\
\hline
\end{tabular}
\end{table*}

\newpage

\begin{table*}[t]
\centering
\caption{[Continuation]}
\label{tab:isotope_abundances}
\begin{tabular}{lcccccc}
\hline\hline
Isotope & $N_{\odot}$ & $N_r$ & $N_r/N_{\odot}$ & $N_s$ & $N_s/N_{\odot}$ & $\Delta\sigma/\langle\sigma\rangle$ \\
 & $(\times 10^6~\mathrm{Si})$ & $(\times 10^6~\mathrm{Si})$ & (\%) & $(\times 10^6~\mathrm{Si})$ & (\%) & (\%) \\
\hline
$^{175}$Lu & $3.73 \times 10^{-2}$ & $3.05(3) \times 10^{-2}$ & 82 & $6.75(31) \times 10^{-3}$ & 18 & 1 \\
$^{176}$Lu & $1.10 \times 10^{-3}$ & $1.04(1) \times 10^{-3}$ & 94 & $6.05(94) \times 10^{-5}$ & 6 & 1 \\
$^{176}$Hf & $8.00 \times 10^{-3}$ & $7.99(22) \times 10^{-3}$ & 100 & $8.00(21549) \times 10^{-6}$ & 0 & 3 \\
$^{177}$Hf & $2.90 \times 10^{-2}$ & $2.40(2) \times 10^{-2}$ & 83 & $5.02(23) \times 10^{-3}$ & 17 & 1 \\
$^{178}$Hf & $4.30 \times 10^{-2}$ & $1.83(4) \times 10^{-2}$ & 42 & $2.47(4) \times 10^{-2}$ & 58 & 1 \\
$^{179}$Hf & $2.10 \times 10^{-2}$ & $1.23(2) \times 10^{-2}$ & 59 & $8.65(18) \times 10^{-3}$ & 41 & 1 \\
$^{180}$Hf & $5.50 \times 10^{-2}$ & $5.83(67) \times 10^{-3}$ & 11 & $4.92(7) \times 10^{-2}$ & 89 & 1 \\
$^{181}$Ta & $2.17 \times 10^{-2}$ & $1.16(4) \times 10^{-2}$ & 53 & $1.01(4) \times 10^{-2}$ & 47 & 2 \\
$^{180}$W & $2.00 \times 10^{-4}$ & $1.91(20) \times 10^{-4}$ & 96 & $9.00(1951) \times 10^{-6}$ & 4 & 10 \\
$^{182}$W & $3.80 \times 10^{-2}$ & $1.38(15) \times 10^{-2}$ & 36 & $2.42(15) \times 10^{-2}$ & 64 & 4 \\
$^{183}$W & $2.10 \times 10^{-2}$ & $7.94(91) \times 10^{-3}$ & 38 & $1.31(9) \times 10^{-2}$ & 62 & 4 \\
$^{184}$W & $4.40 \times 10^{-2}$ & $9.33(529) \times 10^{-3}$ & 21 & $3.47(53) \times 10^{-2}$ & 79 & 12 \\
$^{186}$W & $4.10 \times 10^{-2}$ & $2.36(22) \times 10^{-2}$ & 58 & $1.74(22) \times 10^{-2}$ & 42 & 5 \\
$^{185}$Re & $2.11 \times 10^{-2}$ & $1.54(9) \times 10^{-2}$ & 73 & $5.70(85) \times 10^{-3}$ & 27 & 4 \\
$^{187}$Re & $3.80 \times 10^{-2}$ & $3.44(19) \times 10^{-2}$ & 91 & $3.57(193) \times 10^{-3}$ & 9 & 5 \\
$^{186}$Os & $1.10 \times 10^{-2}$ & $1.06(5) \times 10^{-2}$ & 96 & $3.85(452) \times 10^{-4}$ & 4 & 4 \\
$^{187}$Os & $9.00 \times 10^{-3}$ & $5.63(30) \times 10^{-3}$ & 63 & $3.37(30) \times 10^{-3}$ & 37 & 3 \\
$^{188}$Os & $9.00 \times 10^{-2}$ & $6.46(43) \times 10^{-2}$ & 72 & $2.54(43) \times 10^{-2}$ & 28 & 5 \\
$^{189}$Os & $1.10 \times 10^{-1}$ & $1.05(3) \times 10^{-1}$ & 95 & $5.28(305) \times 10^{-3}$ & 5 & 3 \\
$^{190}$Os & $1.79 \times 10^{-1}$ & $1.53(7) \times 10^{-1}$ & 85 & $2.61(71) \times 10^{-2}$ & 15 & 4 \\
$^{192}$Os & $2.78 \times 10^{-1}$ & $2.69(12) \times 10^{-1}$ & 97 & $9.17(1216) \times 10^{-3}$ & 3 & 4 \\
$^{191}$Ir & $2.32 \times 10^{-1}$ & $2.28(7) \times 10^{-1}$ & 98 & $4.41(739) \times 10^{-3}$ & 2 & 3 \\
$^{193}$Ir & $3.91 \times 10^{-1}$ & $3.86(28) \times 10^{-1}$ & 99 & $5.47(2754) \times 10^{-3}$ & 1 & 7 \\
$^{192}$Pt & $10.00 \times 10^{-3}$ & $1.88(41) \times 10^{-3}$ & 19 & $8.12(41) \times 10^{-3}$ & 81 & 4 \\
$^{194}$Pt & $4.00 \times 10^{-1}$ & $3.80(11) \times 10^{-1}$ & 95 & $1.96(113) \times 10^{-2}$ & 5 & 3 \\
$^{195}$Pt & $4.12 \times 10^{-1}$ & $4.04(12) \times 10^{-1}$ & 98 & $8.24(1216) \times 10^{-3}$ & 2 & 3 \\
$^{196}$Pt & $3.08 \times 10^{-1}$ & $2.70(10) \times 10^{-1}$ & 88 & $3.79(96) \times 10^{-2}$ & 12 & 3 \\
$^{198}$Pt & $8.90 \times 10^{-2}$ & $8.89(38) \times 10^{-2}$ & 100 & $8.90(37872) \times 10^{-5}$ & 0 & 4 \\
$^{197}$Au & $2.01 \times 10^{-1}$ & $1.89(2) \times 10^{-1}$ & 94 & $1.23(23) \times 10^{-2}$ & 6 & 1 \\
$^{198}$Hg & $3.90 \times 10^{-2}$ & $6.71(338) \times 10^{-3}$ & 17 & $3.23(34) \times 10^{-2}$ & 83 & 9 \\
$^{199}$Hg & $6.50 \times 10^{-2}$ & $5.10(40) \times 10^{-2}$ & 78 & $1.40(40) \times 10^{-2}$ & 22 & 6 \\
$^{200}$Hg & $8.90 \times 10^{-2}$ & $4.28(93) \times 10^{-2}$ & 48 & $4.62(93) \times 10^{-2}$ & 52 & 10 \\
$^{201}$Hg & $5.10 \times 10^{-2}$ & $3.07(27) \times 10^{-2}$ & 60 & $2.03(27) \times 10^{-2}$ & 40 & 5 \\
$^{202}$Hg & $1.14 \times 10^{-1}$ & $3.37(61) \times 10^{-2}$ & 30 & $8.03(61) \times 10^{-2}$ & 70 & 5 \\
$^{204}$Hg & $2.60 \times 10^{-2}$ & $2.39(25) \times 10^{-2}$ & 92 & $2.13(248) \times 10^{-3}$ & 8 & 10 \\
$^{203}$Tl & $5.40 \times 10^{-2}$ & $1.04(35) \times 10^{-2}$ & 19 & $4.36(35) \times 10^{-2}$ & 81 & 7 \\
$^{205}$Tl & $1.28 \times 10^{-1}$ & $4.26(95) \times 10^{-2}$ & 33 & $8.54(95) \times 10^{-2}$ & 67 & 7 \\
$^{204}$Pb & $6.60 \times 10^{-2}$ & $8.58(252) \times 10^{-3}$ & 13 & $5.74(25) \times 10^{-2}$ & 87 & 4 \\
$^{206}$Pb & $6.12 \times 10^{-1}$ & $1.66(12) \times 10^{-1}$ & 27 & $4.46(12) \times 10^{-1}$ & 73 & 2 \\
$^{207}$Pb & $6.77 \times 10^{-1}$ & $2.02(39) \times 10^{-1}$ & 30 & $4.75(39) \times 10^{-1}$ & 70 & 6 \\
$^{208}$Pb & $1.94$ & $4.46(2062) \times 10^{-2}$ & 2 & $1.89(21)$ & 98 & 11 \\
$^{209}$Bi & $1.42 \times 10^{-1}$ & $1.13(22) \times 10^{-1}$ & 80 & $2.90(223) \times 10^{-2}$ & 20 & 16 \\
\hline\hline
\end{tabular}
\end{table*}

%
%
\bibliography{bibliography}

\end{document}